\documentstyle[aps,preprint,epsfig]{revtex}

\begin{document}

\title{New measurement and analysis of the $^{7}$Be(p,$\gamma$)$^{8}$B cross
section}

\author{F. Hammache$^{1)}$, G. Bogaert$^{1)}$, P. Aguer$^{1)}$, C.
Angulo$^{1,*)}$, S. Barhoumi$^{4)}$, L.
Brillard$^{3)}$, J.F.~Chemin$^{2)}$, G. Claverie$^{2)}$, A. Coc$^{1)}$, M.
Hussonnois$^{3)}$, M.Jacotin$^{1)}$, J. Kiener$^{1)}$, A.~Lefebvre$^{1)}$, 
J.N. Scheurer$^{2)}$,
J.P. Thibaud$^{1)}$, E. Virassamyna\"{\i}ken$^{2)}$}

\address{
$^{1)}$ CSNSM, IN2P3-CNRS, 91405 Orsay, France\\
$^{2)}$ CENBG, IN2P3-CNRS, et Universit\'{e} de Bordeaux I, 33175 Gradignan,
France\\
$^{3)}$ IPN,IN2P3-CNRS et Universit\'{e} de Paris-Sud, 91406 Orsay, France\\
$^{4)}$ Institut de Physique, USTHB,B.P. 32, El-Alia, Bab Ezzouar, Algiers,
Algeria}

\maketitle

\begin{abstract}

Cross sections for the $^{7}$Be(p,$\gamma$)$^{8}$B reaction have been measured for
E$_{c.m.}$  =
0.35-1.4 MeV using radioactive $^{7}$Be targets. Two independent measurements
carried out with different beam conditions, different targets and detectors
are in excellent agreement. A statistical comparison of these measurements
with previous results leads to a restricted set of consistent data. The
deduced zero-energy S-factor S(0) is found to be 15-20\% smaller than
the previously recommended value. This implies a $^{8}$B solar neutrino flux lower
than previously predicted in various standard solar models.

\end{abstract}


\narrowtext
The $^{8}$B produced in the solar interior via the reaction
$^{7}$Be(p,$\gamma$)$^{8}$B
is the major (or unique) source of high energy neutrinos detected in many
solar-neutrino experiments now operating or in development (Homestake,
Kamiokande, Super-Kamiokande, SNO...[1]). The observed deficit of $^{8}$B solar
neutrinos when compared to the predictions of solar models [1,2] might have
its origin, at least partly, in the value of the $^{7}$Be(p,$\gamma$)$^{8}$B
cross section at
very low energy ($\sim$ 20 keV) since the magnitude of the $^{8}$B solar
neutrino flux is directly proportional to the rate of this reaction.
Moreover, the interpretation of the various experiments in terms of neutrino
oscillations depends on the reliability of the measured cross sections. For
instance, it has been argued [3] that the prediction for the charged to
neutral current ratio in SNO is strongly dependent on the estimation of the
$^{8}$B neutrino flux. 

There are six direct measurements of the $^{7}$Be(p,$\gamma$)$^{8}$B cross
section [4-9] using radioactive $^{7}$Be targets and proton beams, the most
recent dating back to 1983. In addition, a result [10] was obtained in 1994
studying the Coulomb dissociation of $^{8}$B at 50 MeV/u energy. The four most
precise measurements [5-7,9] are grouped in two distinct pairs which are
in agreement with regard to the energy dependence but in disagreement with
regard to the absolute value. Zero-energy S-factors (S(E$_{c.m.}$) =
$\sigma$(E$_{c.m.}$)E$_{c.m.}$e$^{-2\pi\eta}$,
and $\eta$ = e$^{2}$Z$_{1}$Z$_{2}$/$\hbar$v) S(0) are deduced from measurements
by an extrapolation based on theoretical calculations of the energy
dependence of the cross section. The resulting S(0) are found to disagree by
as much as 40 \%, making this quantity the most uncertain input to solar
models. Therefore, it appears highly desirable to perform new measurements
of the $^{7}$Be(p,$\gamma$)$^{8}$B cross section. 

In this letter, we report measurements of the $^{7}$Be(p,$\gamma$)$^{8}$B
cross section for 0.35 MeV $\le$E$_{c.m.}\le$ 1.4 MeV using
radioactive $^{7}$Be targets. Special
attention
was devoted to checking the
internal consistency of the measurements and to reducing the uncertainties
with the aim of restricting the available data for
$^{7}$Be(p,$\gamma$)$^{8}$B to a set of consistent measurements. 

The experiment was performed at the Bordeaux 4 MV Van de Graaff accelerator.
The target, produced via the $^{7}$Li(p,n)$^{7}$Be reaction using the same accelerator,
consisted of $^{7}$Be oxide deposited on Pt disk. 
Details of the target preparation will appear elsewhere [11]. The
$^{7}$Be(p,$\gamma$)$^{8}$B
cross section was measured by detecting the delayed $\alpha$  particles
following the $\beta^{+}$
decay of $^{8}$B. The bombard-count cycle was as follows : the target was
irradiated for 1.54 s with the detectors protected against the flux of backscattered
protons by a metallic iris diaphragm. The beam was then deflected off the
target (transit time = 0.24 s) via an electrostatic device for 1.52 s. During
this phase, a mechanical shutter stopped neutral hydrogen. The iris
diaphragm was then opened and a time window of 1.34~s was defined for $\alpha$
counting before going back to the irradiation position (transit
time = 0.21~s). The target was fixed so that it could be efficiently water-cooled which
was not the case in the previous experiments [5-9] where a rotating arm was
used to transfer the target from the bombarding chamber to the counting
chamber. In consequence, we were able to use currents of typically 25 $\mu$A 
without noticeable degradation of the target. A liquid nitrogen cooled
copper plate was positioned very close to the target to reduce carbon
build-up. The beam was collimated to a spot of approximately 4 by 4 mm at
the target by passing through two diaphragms (8 and 6 mm in diameter) 1.5 m
apart. In addition a third insulated collimator (7 mm diameter) was placed 1
cm in front of the target. The negligible currents measured in all runs on
this collimator gave evidence for the absence of significant instability in
the beam position at the target during a run. The data were recorded event by 
event. Due
to the low data acquisition rate, special precautions were taken against
spurious events using a veto signal which inhibited the acquisition when
an extra detector located outside the reaction chamber was triggered by a
rare electrical noise signal. Moreover in the data analysis, events in which
more than one detector fired were rejected. Beam currents on all collimators
and on the target were measured, digitized and recorded on a computer system
for off-line analysis. To suppress secondary electron emission the large
insulated copper plate acting as LN$_{2}$  cold trap in front of the target and
the last collimator were biased at -300 V. In addition, the beam current was
measured in a Faraday cup before and after each run and found to be in good
agreement with measurements on the target to within 2\%. 

Two independent measurements were carried out. For the first run, referred
to as (95), the target activity was 10.4 $\pm$ 0.4 mCi and the detector consisted
of a set of four passivated implanted silicon counters, with a total active surface of
12 cm$^{2}$ and a 100 $\mu$m depletion depth. For the second experiment, referred to
as (96), the target activity was increased to 26.9 $\pm$ 0.5 mCi and four
surface barrier detectors 30 $\mu$m thick were used. With this improved set-up,
cross sections were measured at ten energies (E$_{c.m.}$) ranging from 0.35
to 1.4 MeV. Only comments on the analysis of (96) are given here. The
analysis of (95) was very similar with however slightly larger error bars
mainly due to the deconvolution process of the $\alpha$ spectra. Cross sections
were obtained from the integrated $\alpha$ particle yields in a manner similar to
that described in Ref. [9]. Two typical spectra of delayed $\alpha$
particles taken at different energies are
shown in fig. 1. The small thickness of the detector and its segmentation
into four sectors strongly reduced the pile-up events seen as a dashed steep 
line extending up to 0.760 MeV in the figure and due to photoelectrons
created by the 478 keV $\gamma$ rays. In deducing cross sections, counts in the range
from 0.76 to 5 MeV were integrated and a small correction factor for
energy cutoffs (typically 1.05 $\pm$ 0.01) was calculated from a curve fitted to
the data in the same energy range. This curve was deduced from the
actual  $\alpha$ spectrum shape given in Ref. [12] after correction of energy
loss and energy straggling of the emitted $\alpha$ particles (fluctuations
in the range of the recoiling $^{8}$B were also considered).
As shown in fig. 1, very good agreement is obtained without introducing any
free parameters into the analysis except the normalization factor. It was
checked that the same corrected number of counts within statistical error
bars, was obtained when varying the value of the low-energy cut. The same
procedure applied equally well to $^{7}$Li(d,p)$^{8}$Li (see below). The background
was determined in an extensive series of measurements alternating between
beam-off and beam-on. It contributed from $\le$ 2\% of the $\alpha$ yield at
E$_{c.m.}$ $\ge$ 0.5 MeV to $\sim$ 7\% at the lowest energies. In addition, the
background due to a possible deuteron contamination of the H$^{+}$ beam was found
to be less than 0.1 \% at all energies. Effective reaction energies were
determined from measurements of the target thickness (4.0 $\pm$ 0.4 keV at Ep =
441 keV) and of the carbon build-up by consistent RBS measurements and (d,p)
reaction analysis of $^{12}$C, $^{16}$O performed many times during the course of the
experiment. The overall corrections for target thickness and carbon build-up
lead to an effective energy uncertainty of less than 0.3\%. The
beam energy was calibrated to $\pm$ 0.1\% from thick target yield curves at
resonances in the $^{19}$F(p,$\alpha$$\gamma$)$^{16}$O
[$^{27}$Al(p,$\gamma$)$^{28}$Si] reaction at proton energies of 340.46 keV
and 871.11 keV [632.6 keV, 991.8 keV]. 

The product of
initial $^{7}$Be areal density and efficiency of the $\alpha$ detector,
N$_{^{7}Be}$(0) x $\epsilon$, was
measured by two methods as initiated in Ref. [9]. In the first method, the
total activity of $^{7}$Be was determined several times by measuring the yield of
the 478 keV $\gamma$ ray with a Ge detector and using the known branching ratio of
10.53 $\pm$ 0.036 \% for the electron capture of $^{7}$Be to the first excited state of
$^{7}$Li [13]. The detector efficiency was obtained using standard $\gamma$ ray sources
calibrated to within 1\% uncertainty. After fitting the $^{7}$Be decay function
to the various measurements ($\chi^{2}$ = 0.43), we found an initial total activity
of 26.9 $\pm$ 0.5 mCi. For the whole duration of the experiment no loss of
activity due to beam impact was observed as indicated by the excellent fit
to the data. The target surface was measured by computer scanning of a
photographic enlargement of the target where the $^{7}$Be deposit clearly
appears. Furthermore, $\gamma$-activity scanning of the target was performed with a
Ge detector collimated with a 0.85 mm diameter aperture in a 15 cm thick
lead absorber. This measurement gave the degree of target uniformity of the
$^{7}$Be density and a total target area which was consistent with the 
previous one (S =0.47 $\pm$ 0.02 cm$^{2}$). The beam position at the target 
was systematically
determined before and after each run and found stable at each energy. The
$^{7}$Be areal density at the target spot was finally determined
($\pm$ 5 \% uncertainty) run by run by averaging the results
of the $\gamma$-ray scan over the beam spot dimensions and
normalizing to the total activity per surface unit. An extensive and
consistent series of measurements was made to determine the efficiency
$\epsilon$  of
the $\alpha$ detector, using calibrated $^{241}$Am sources of different diameters and
different centerings deposited onto Pt backings identical to those used in the
experiment and with the same source-detector geometry. We found $\epsilon$ =
0.107 $\pm$ 0.002. In the second method, N$_{^{7}Be}$(t) x $\epsilon$ was
independently determined with the
same experimental set-up from the delayed $\alpha$ yield of the reaction
$^{7}$Li(d,p)$^{8}$Li. Averaging over five measurements of this reaction yielded a
value for N$_{^{7}Be}$(0) x $\epsilon$
very close to the same quantity as measured directly.
Specifically, the ratio is 1.01 $\pm$ 0.08 using a value of 147 $\pm$ 11 mb [14] for
$^{7}$Li(d,p)$^{8}$Li at the 0.61 MeV resonance. Hence, both methods gave 
identical results for the cross section with however lower error bars for 
the first one owing to the extensive and consistent series of measurements 
devoted to obtaining the detector efficiency and the target activity at the 
beam spot, as explained above.   

Results in the form of astrophysical S-factors are given in fig. 2. (see
also fig. 3). No measurements were
carried out in the region of the resonance at E$_{c.m.}$= 0.660 MeV which has no
significant contribution to the cross section in the energy range E$_{c.m.}$=
0-0.5 MeV and E$_{c.m.}$= 0.85-1.4 MeV where our measurements were concentrated.
In that region, the E1 direct capture process is largely dominant. At E$_{c.m.}$=
0.88 MeV, four independent proton bombardments were made, three [one] of
them with the experimental set-up (96) [(95)]. The four experiments were
found to be in excellent agreement with a reduced $\chi^{2}$= 1.1. The same
excellent agreement was observed for two independent measurements at E$_{c.m.}$ =
0.497 MeV. The consistency of the whole set of independent
measurements made with different beam conditions, different targets and
detectors strongly supports the reliability of the data (and the correct
evaluation of the uncertainties) and experimental bias negligible
compared to the quoted errors. 

A comparison of our measurements with existing data is shown in fig. 2.
In a previous analysis, Johnson et al. [15] used the
results of Refs. [5-7,9] in the averaging process for determining S(0)
despite the fairly large spreading of the data. However, the present work
provides an additional strong constraint on the consistency of the various
experiments (see fig. 2).
To be quantitative, we have performed a $\chi^{2}$ test on the S(0)
deduced by a least squares normalization of
the same S(E) curve calculated by Descouvemont et al. [16] to each of
the data sets considered. The used experimental values were in
the energy range from 0.11 to 0.5 MeV and 0.87 to 1.4 MeV in which
the resonance contributes no more than 3.4\% to the data (the corresponding
small contributions were substracted using results of Ref. [9]). Such a fit
is shown in fig. 3 for our data.
Note that the fits were performed for each experiment using relative error
bars. The resulting uncertainty in S(0) was
then combined in quadrature with ''systematic'' uncertainties applied on the same
footing to every energy point of a given experiment.
The obtained S(0) and associated
error bars are given in table 1. Since most of the experiments rely on
normalization to $^{7}$Li content in target via the $^{7}$Li(d,p)$^{8}$Li 
cross section, we applied the $\chi^{2}$ test to the S(0) corresponding to
such analyses for all experiments including our own and that of Filippone. 
As we used the same value $\sigma$$_{dp}$ = 147 $\pm$ 11 mb [14] for the normalization of all the
experiments, the contribution to the uncertainty related
to $\sigma$$_{dp}$ was not included the error bars for the $\chi^{2}$ test. 
On this basis, the consistency of the five sets of data is ruled out at 99.9\% C.L.
By way of precaution, we checked that the above conclusion does not depend
closely on the estimation of the error bars.
Specifically, when increasing the uncertainties by a
factor of two (three) the consistency of the data is still ruled out at
99.5\% (95\%)C.L .
On the contrary, a complete consistency is found (reduced $\chi^{2}$= 0.5) with the
actual errors when considering only our data and those of Filippone and
Vaughn.
The above analysis is independent of the fitted curve  so long  as the fits 
are good for all sets of data. The latter point is clearly verified as shown 
by the obtained reduced $\chi^{2}$ given in table 1. 

The consistent S(0) values of this work and of Filippone and
Vaughn have been averaged taking into account that some of the experiments
were normalized to the same value of $\sigma$$_{dp}$ (the uncertainty in
$\sigma$$_{dp}$ was then treated as an overall systematic uncertainty). For
our experiment we took the uncertainty in S(0) quoted in table 1 which arises
from the normalization procedure via direct measurement of $^{7}$Be activity.
The final result is $<$S(0)$>$ = 18.3 $\pm$ 0.8 eV b, very close to the
value of 18.5 $\pm$ 1.0 obtained with our data alone.
A similar averaged value of 18.5 $\pm$ 1.0 eV b is found when the fits are
restricted to the maximum energy of 0.5 MeV using our data and those of
Ref. [9] (note that the goodness of individual fits to each set of data is
found to be excellent whatever the energy range considered [17]). The same
analysis, when
performed with the curve calculated by Johnson et al. [15], leads to a value of
18.3 $\pm$ 1.0 eV b. 

The present value for $<$S(0)$>$ is significantly lower than
the previously recommended value of 22.4 $\pm$ 2.1 eV b given by 
Johnson et al. [15].
The reason is essentially that we did not consider the results of Refs. [5,6] 
in our averaging process in contrast to Johnson et al. (an additional reduction
arises from the different values adopted for $\sigma$$_{dp}$). 

Finally, the obtained $<$S(0)$>$ value implies a significant reduction of
15-20\% in the $^{8}$B solar neutrino flux. We are presently developing
experiments at lower energies to further reduce the overall uncertainty on
the zero energy S-factors for this reaction. 

We gratefully acknowledge discussions with P. Descouvemont, M. Harston, Y.
Llabador and S. Turck-Chieze. We thank J.J. Correia for experimental support
and L. Lavergne (IPN-Orsay) for providing us with the 30 $\mu$m thick detectors.
This work was supported in part by Region Aquitaine and the E.C. under the
HCM network contract 930339.
\begin{figure}
\begin{center}
\mbox{\epsfig{file=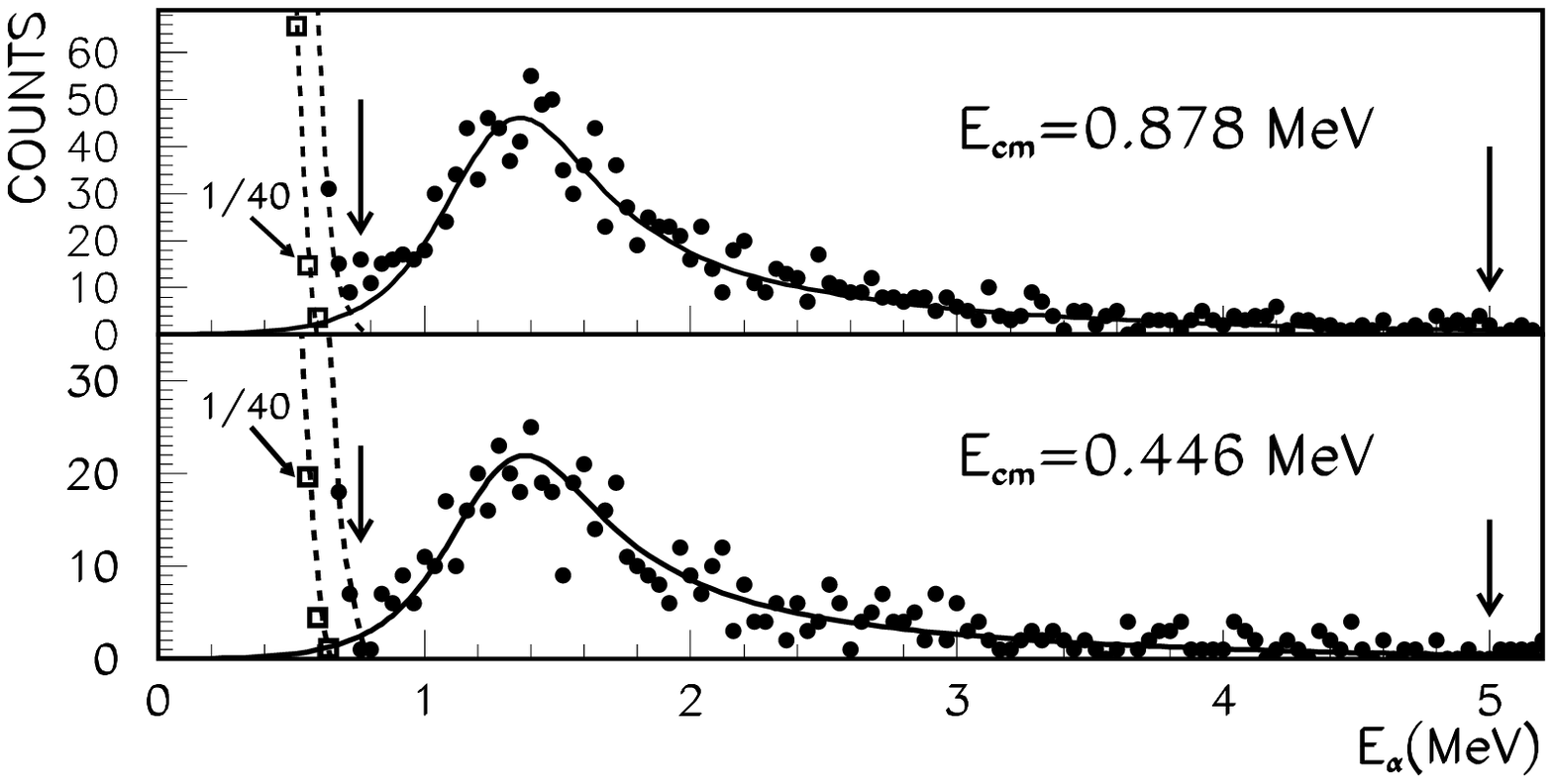,width=16cm}}
\vspace{-0.4 cm}
\caption{Delayed $\alpha$-particle spectrum from decay of $^{8}$B at two different
energies. The $\alpha$ particle yields were integrated from the energy cutoffs
indicated by the arrows. The solid curve is a fit to the data as explained
in the text. The dashed line is a fit to the low-energy background due to
pile-up events. For squares, the y scale is divided by 40.}
\end{center}
\vspace{-0.3 cm}
\end{figure}

\newpage

\begin{figure}
\vspace{-0.5 cm}
\begin{center}
\mbox{\epsfig{file=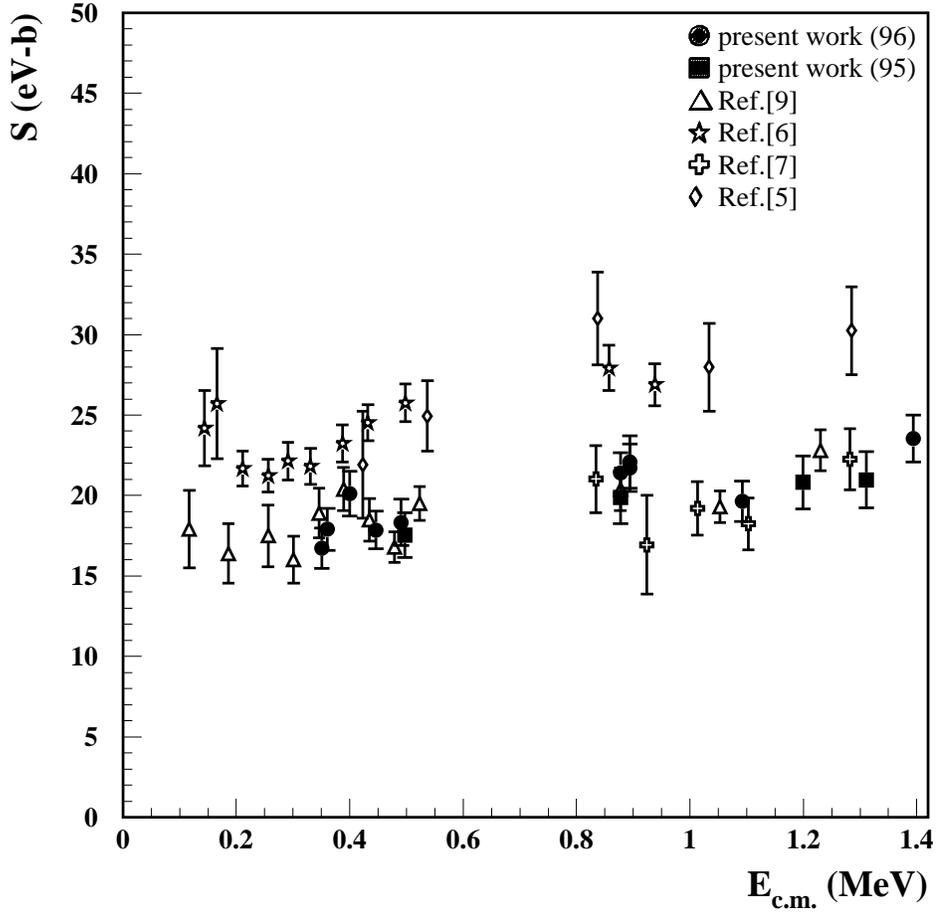,width=14cm}}
\caption{The $^{7}$Be(p,$\gamma$)$^{8}$B S factors obtained from our two series of
measurements together with existing data shown at energies outside the M1
resonance. The data were renormalized using $\sigma$ = 147
$\pm$ 11 mb [14] for $^{7}$Li(d,p)$^{8}$Li at E$_{c.m.}$ = 0.61 MeV. The
error bars represent only the relative uncertainties in the points.}
\end{center}
\end{figure}

\newpage

\begin{figure}
\begin{center}
\mbox{\epsfig{file=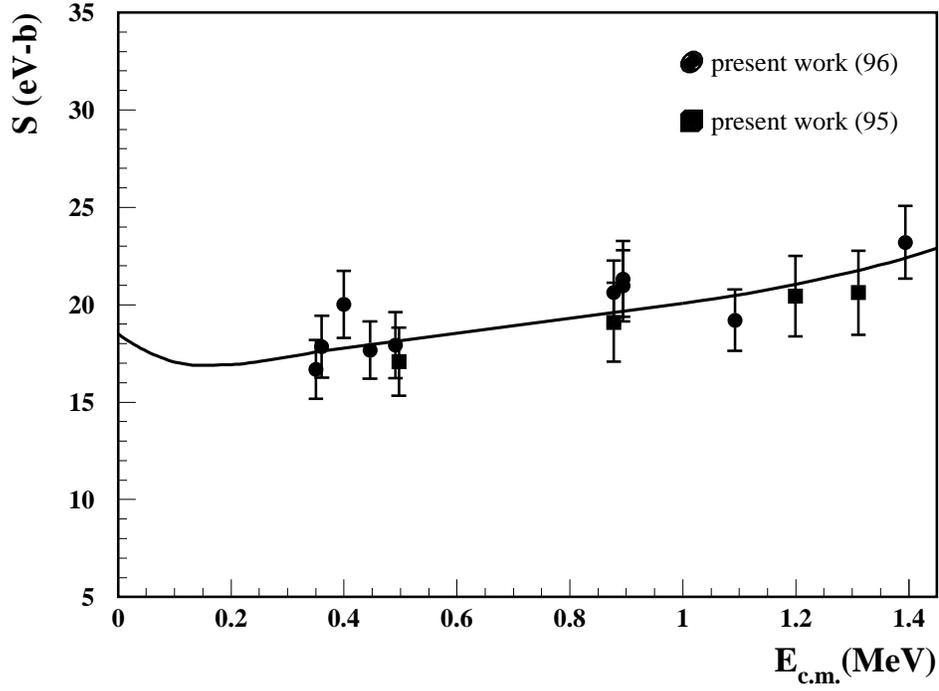,width=14cm}}
\caption{S-factors from the present work and typical fit using theoretical
curve of Descouvemont et al. [16] for the non resonant capture (their
resonant M1 contribution has been substracted).The only free parameter in 
the fit is a normalization factor. Overall error bars corresponding to the
first method of analysis (see text) are drawn.}
\end{center}
\end{figure}

\newpage
\begin{center}
\begin{table}[htb]
\caption{Extrapolation of the S-factors to zero energy using the
energy dependence of Descouvemont et al. [16]. All the experiments labelled
a) were normalized with $\sigma$$_{dp}$ = 147 $\pm$ 11 mbarns for
$^{7}$Li(d,p)$^{8}$Li. The uncertainties given in
the second column are overall uncertainties (see text)). The uncertainty in
$\sigma$$_{dp}$ must be substracted when comparing the experiments labelled
a).}
\hspace{0.5cm}
\begin{tabular}{lccc}
 Experiment  & S(0) & reduced $\chi^{2}$ \\
\hline
Ref. [5] $^{a)}$ & 25.8 $\pm$ 2.2 & 0.55 \\

Ref. [6] $^{a)}$ & 24.3 $\pm$ 2.0  & 0.74 \\

Ref. [7] $^{a)}$ & 17.4 $\pm$ 1.6 & 0.75 \\

Ref. [9] $^{a)}$ & 18.4 $\pm$ 2.2$^{c)}$ & 1.1 \\

Ref. [9] $^{b)}$ & 18.4 $\pm$ 2.4$^{c)}$ & 1.1 \\

Present $^{a)}$ & 18.5 $\pm$ 1.7 & 0.65 \\

Present $^{b)}$ & 18.5 $\pm$ 1.0 & 0.65 \\
\end{tabular}
\vspace{0.5 cm}
a) Cross section determined from the $^{7}$Li(d,p)$^{8}$Li cross section. 
b) Cross section determined from $\gamma$-ray activity. 
c) Error bar deduced from that given in Ref. [9] assuming that the random
error arising from the fit is similar in Ref. [9] and in the present analysis.
\end{table}
\end{center}

\begin{thebibliography}{20}
\bibitem{ } It is worth noting that no M1 contribution at E$_{c.m.}$ $>$ 1
MeV has to be invoked to obtain a good agreement between experimental data and
the calculated curve of Ref. [16] in contrast to a previous analysis [7]
which used energy dependence for which only s-wave protons were included and
d-waves were neglected.
\end{thebibliography}
\end{document}